\pdfoutput=1

\documentclass[twocolumn,nofootinbib,showpacs,preprintnumbers,amsmath,amssymb]{revtex4}
\usepackage{graphicx}% Include figure files \usepackage{dcolumn}% Align table columns on decimal point \usepackage{bm}% bold math
\usepackage{epsfig}
\newcommand{\be}{\begin{equation}}
\newcommand{\ee}{\end{equation}}
\newcommand{\ba}{\begin{eqnarray}}
\newcommand{\ea}{\end{eqnarray}}

\newcommand{\GeV}{~\mathrm{GeV}}

\newcommand{\kahler}{K\"{a}hler }
\begin{document}
\preprint{MCTP-09-13} \preprint{UCB-PTH-09/11}
\title{CP-violating Phases in M-theory and Implications for EDMs}% Force line breaks with
\author{Gordon Kane$^{1}$}
\author{Piyush Kumar$^2$}
\author{Jing Shao$^1$}
\affiliation{$^1$Michigan Center for Theoretical Physics, Ann Arbor, MI 48109 USA \\ \\
$^2$Department of Physics, University of California,
Berkeley, CA 94720 USA \;\;$\&$\\Theoretical Physics Group, Lawrence Berkeley National Laboratory, Berkeley, CA 94720}
\date{\today}% It is always \today, today

\begin{abstract}
We demonstrate that in effective theories arising from a class of ${\cal
N}=1$ fluxless compactifications of M-theory on a $G_2$ manifold
with low energy supersymmetry, CP-violating phases do not appear in the
soft-breaking Lagrangian except via the Yukawas appearing in the trilinear parameters.
Such a mechanism may be present in other string compactifications as well; we
describe properties sufficient for this to occur. CP violation is generated via the Yukawas since the soft trilinear
matrices are generically not proportional to the Yukawa matrices.  Within the
framework considered, the estimated theoretical upper bounds
for electric dipole moments (EDM) of the electron, neutron and
mercury are all within the current experimental limits and could
be probed in the near future.
\end{abstract}
%\pacs{11.25.Mj 11.25.Wx 11.25.Yb 12.10.-g 12.60.Jv 14.80.Ly}% PACS, the Physics and Astronomy
\maketitle
\newpage
\vspace{-1.2cm}
%\tableofcontents

\section{Introduction}

The null measurements of the electric dipole moments (EDMs) of the
neutron \cite{Baker:2006ts}, and recently, heavy atoms like
Thallium ($^{205}$Tl)\cite{Harris:1999jx,Regan:2002ta} and Mercury
($^{199}$Hg) \cite{Romalis:2000mg,Griffith:2009zz}, have put very
strong constraints on the amount of CP violation from new physics
beyond the Standard Model (SM). The precision of these
measurements is expected to significantly improve in a few years.
If an excess above the SM prediction is observed, it requires the
presence of new physics beyond the SM. However, since the EDMs,
even if observed, are already \textquotedblleft small", this
strongly suggests that the new physics must be such that it has an
underlying mechanism to naturally suppress EDMs.

In general versions of supersymmetric extensions of the Standard
Model, new sources of CP violation can arise from complex phases
of the soft supersymmetry breaking parameters. These phases are
therefore tightly constrained to be small
\cite{Masiero:1997bn,Abel:2001vy}(or to have cancellation
\cite{Ibrahim:1997nc,Ibrahim:1997gj,Ibrahim:1998je,Brhlik:1998zn})
for TeV scale superpartners. Thus, from a theoretical perspective,
the existence of such small phases has to be explained by some
underlying mechanism. Many studies of supersymmetric models from a
low-energy phenomenological perspective focus on the
\emph{mediation} mechanism and only parameterize the supersymmetry
breaking. Explaining small soft CP-violating phases, which
requires a dynamical understanding of supersymmetry breaking, is
especially challenging as this is not available in such a
framework. Without a specification of the supersymmetry breaking
mechanism, this problem exists in both gravity and gauge-mediated
models of supersymmetry breaking in general.

Put differently, whenever supersymmetry is treated as a general
TeV-scale effective theory both the values and phases of the
soft-breaking masses are treated as arbitrary, and EDMs are
typically much larger than experimental values. Many people have
argued that such large EDMs are implied or required from
supersymmetry, and that this is a problem for supersymmetry. Such
arguments ignore the fact that any underlying theory will predict
and relate phases. This implies that the underlying theory of
which low energy supersymmetry is a low energy limit has a
structure that suppresses or relates the low scale phases.

Substantial progress has been made towards understanding dynamical
supersymmetry breaking, especially in recent years. In this work,
we will be interested in dynamical mechanisms of supersymmetry
breaking with low superpartner masses which can be naturally
embedded in the framework of an underlying microscopic theory like
String/M-theory. In particular, we study the effective
four-dimensional theory resulting from fluxless $\mathcal{N}=1$
compactifications of M-theory with chiral
matter~\cite{Acharya:2006ia}. These are especially interesting
because a hierarchy between the Electroweak and Planck scale is
generated, and all geometric moduli are stabilized, at the same
time~\cite{Acharya:2007rc, Acharya:2008hi}. We find that the supersymmetry breaking
and mediation dynamics is such that it naturally gives rise to vanishing
CP-violating phases from supersymmetry breaking at leading order,
providing an excellent starting point to explain
suppressed EDMs. The mechanism is a non-trivial generalization of
an old idea~\cite{Choi:1993yd} (and more
recently~\cite{Conlon:2007dw}), and may also apply to other
classes of string compactifications where moduli are stabilized in
a de Sitter vacuum.

Although the CP-violating phases from supersymmetry breaking vanish
at leading order, there could still be significant
contributions to CP violation in the flavor-diagonal sector
in principle. First, in the M-theory
framework, the trilinear matrices are typically not proportional
to the Yukawa matrices after moduli are stabilized, which in
general leads to non-trivial CP-violating phases in the trilinear
A-terms in the basis of quark and lepton mass eigenstates and therefore
generates non-zero EDMs \cite{Abel:2001vy,Abel:2001cv}. The
estimated upper bounds on EDMs are all within the current
experimental limits. For some values of parameters, some upper
bounds on the EDMs are close to the experimental limits. As will be clear, two
features - large sfermion masses and trilinears, and hierarchical Yukawa
textures, both natural within the M-theory framework, are important
for getting viable but interesting EDMs results. In addition,
we argue that even though higher order corrections to the \kahler
potential exist, they do not give rise to new CP-violating phases.
Finally, it should be remarked that the EDM results are robust
anytime the trilinears dominantly acquire their phases from Yukawas
and the mass spectrum is as dictated by the M-theory framework.

It is worth mentioning that the solution to the supersymmetric CP
problem here is largely independent of any particular solution to
flavor issues as long as they satisfy a certain criterion. As will
be seen, the only feature of the flavor structure used in
computing results for EDMs is that the sfermion mass matrices for visible matter in
the super-CKM basis are approximately
flavor diagonal at low energies. Therefore, the results hold true
for any proposed solution to the flavor problem which is
consistent with the above feature. For the M-theory
framework in particular, this approximately flavor diagonal structure
arises due to the presence of $U(1)$ symmetries under which the chiral
matter fields are charged\cite{Bourjaily:2009ci}. The spontaneous breaking of
these symmetries may introduce small non-diagonal components as
long as it occurs at a scale sufficiently below the Planck scale. In the
M-theory framework, large sfermion masses $\gtrsim 10$ TeV
already mitigate the FCNC problems. In addition, small off-diagonal components
(after going to the super-CKM basis) suppressed by
an order of magnitude or more could arise from the approximate flavor-diagonality
of the \kahler metric mentioned above and/or due to family symmetries which could
be present in the underlying theory. This would then probably be consistent with
all FCNC observables. This paper focusses on CP
violation arising in the flavor-diagonal sector, which is present in general even if FCNC problems,
arising from off-diagonal terms in the squark mass matrices in the super-CKM basis, are solved; hence
it is largely decoupled from flavor physics. A more detailed discussion of flavor issues will
appear elsewhere.

The plan of the paper is as follows. In section \ref{leading}, we
review the basic mechanism of supersymmetry breaking and its
implications for soft CP-violating phases at leading order.
%An important point to note is that the soft phases vanish at leading
%order even for dS vacua with a (tiny) positive cosmological constant.
In section \ref{textures}, we discuss the connection
between yukawa textures and imaginary parts of the diagonal
trilinear matrix components. After a brief discussion of EDMs
using the low energy effective Lagrangian and the present
experimental limits in section \ref{edm}, we compute the detailed
predictions for EDMs within this framework in section
\ref{predict}. In section \ref{general}, we discuss the effect of 
possible higher order corrections to the superpotential and K\"{a}hler potential 
in this class of compactifications and argue that the results obtained are robust against these 
corrections. We conclude in section \ref{conclude}. The
appendices deal with some technical details of the computations.

\section{Small CP-violating Phases from SUSY breaking}\label{leading}

In fluxless compactfications of M-theory\cite{Acharya:2007rc}, the moduli
superpotential is entirely generated non-perturbatively, and
hence, exponentially suppressed relative to the Planck scale. This
is crucial for both stabilizing the moduli $z_i$ and generating
the hierarchy. The strong gauge dynamics resides in a
three-dimensional submanifold of the internal manifold which
generically does not intersect the three-dimensional submanifold
where the supersymmetric standard model particles live as these
three-manifolds are embedded inside a seven dimensional internal
manifold. For simplicity, we consider two non-abelian
asymptotically free gauge groups with at least one of them assumed
to contain light charged matter fields ${\cal Q}$ and ${\cal
\tilde{Q}}$ (with $N_f < N_c$), and an associated meson $\phi =
(\tilde {\cal Q} {\cal Q}^{T})^{1/2}$ in the low energy. The
strong gauge dynamics in the hidden sector stabilizes the moduli
of the $G_2$ manifold, and dynamically generates a supersymmetry
breaking scale with $\mathcal{O}$(10) TeV gravitino mass. The supersymmetry breaking
is then mediated to the visible sector through gravitational
($m_{p}$ suppressed) interactions.

\subsection{Superpotential and \kahler Potential}\label{kahlersuper}
To be self-contained, in the following we briefly discuss the effective action
of the fluxless compacifications of M-theory studied in detail in \cite{Acharya:2007rc}. We will emphasize some important features
that are crucial for our results. Further details can be found in above references.

First, the superpotential can be separated into two parts:
\begin{eqnarray}\label{superW}
  W= \hat W + Y^{\prime}_{\alpha\beta\gamma} C^{\alpha}C^{\beta}C^{\gamma}
\end{eqnarray}
where $\hat W$ depends only on the moduli $z_i=s_i + i t_i$ and
the meson $\phi$. Here $C_{\alpha}$ are the matter fields in the
minimal supersymmetric standard model (MSSM) with $\alpha$ being
higgs, quark or lepton chiral superfields.
$Y^{\prime}_{\alpha\beta\gamma}$ denote the superpotential Yukawa
couplings. The effective Yukawa couplings (still not fully
normalized) in the MSSM are given by $Y_{\alpha\beta\gamma}=
e^{K/2}Y^{\prime}_{\alpha\beta\gamma}$. The connection to the
usual convention in the MSSM can be made by taking the first index to
be the Higgs fields, the second to be the quark doublets, and the
third to be the quark singlets, for example, $Y_{H_u Q_i
u_j}\equiv Y_{ij}^{u}$. In the M-theory framework, an elegant way
to generate Yukawa couplings $Y^{\prime}_{\alpha\beta\gamma}$ is from
membrane instantons \cite{Atiyah:2001qf, Bourjaily:2009vf}, which also
depend holomorphically on the moduli $z_i$ in general as they measure
the volume of the manifold which the instanton wraps. It appears natural
to generate a hierarchical Yukawa texture from such effects.

The first term $\hat W$ in (\ref{superW}) is the moduli superpotential, and
is generated non-perturbatively from gaugino condensation in two hidden sectors, one of which
is assumed to also contain matter fields (with $N_f < N_c$)
\cite{Affleck:1983mk}:
\begin{eqnarray}\label{superpotential}
\hat W = A_1 \, ({\rm det}(\phi^2))^{a/2}\,e^{-b_1\,f}+A_2
\, e^{-b_2\,f}
\end{eqnarray}
Here $b_{1,2}$ are the beta function coefficients of the two
hidden sector gauge groups and $f$ are the corresponding
gauge kinetic functions given by $f=\sum_{i=1}^{N}\,N_{i} z_{i}$. $\phi = (\tilde
{\cal Q} {\cal Q}^{T})^{1/2}$ denotes the meson fields, and we have suppressed
the flavor indices for simplicity. The parameter $a$ in the superpotential is a rational number -$a=-2/(N_c-N_f)$.

The \kahler potential can be written as
\begin{eqnarray}
K= \hat K + \tilde K_{\alpha\beta} {C^{\alpha}}^{\dagger}C^{\beta}
+ \left(Z_{\alpha\beta}C^{\alpha}C^{\beta} + h.c.\right)
\end{eqnarray}
Here $\hat K$ is the moduli \kahler potential and $\tilde
K_{\alpha\beta}$ is the \kahler metric of matter fields
$C^{\alpha}$. $Z_{\alpha\beta}$ is expected to be non-zero only
for Higgs field $H_{u,d}$, which is needed to generate $\mu$ and
$B$ terms. In these compactifications, charged chiral matter
fields with different flavors are localized at \emph{isolated}
conical singularities \cite{Friedmann:2002ty}. These charged
matter fields, in addition to being charged under the relevant
non-abelian gauge group, are also charged under $U(1)$ factors
which arise from the Kaluza-Klein reduction of the three-form in
eleven dimensional supergravity on two-cycles present in the
internal manifold \cite{Acharya:2004qe}. These $U(1)$'s survive at
low energies as good symmetries to all orders in perturbation
theory and hence must be respected (up to exponentially suppressed
non-perturbative effects). Importantly, it turns out that conical
singularities associated to different flavors cannot carry the
same charges under the $U(1)$'s in a given basis
\cite{Bourjaily:2009ci} (at least in local models). This forbids
the existence of off-diagonal terms in the \kahler potential of the
form $C^{\alpha\,\dag}C^{\beta}$, $\alpha \neq \beta$. Off-diagonal
components may be introduced if these symmetries are spontaneously broken,
but these will be suppressed as long as this occurs sufficiently
(an order of magnitude or more) below the Planck scale.

Thus, the \kahler metric is expected to be approximately flavor diagonal, i.e.
$\tilde K_{\alpha\beta} \approx \tilde K_{\alpha} \delta_{\alpha\beta}$
at the high scale. As argued in \cite{Acharya:2008hi}, the \kahler
potential for localized matter fields $C_{\alpha}$ in the 11D frame
is canonical, i.e. $C_{\alpha}^{\dag}C_{\alpha}$ due to the
absence of local moduli. Going to the Einstein frame implies that
there is an overall dependence on the internal volume $V_X$, but
this still preserves the approximate diagonality. A flavor-diagonal \kahler
metric will lead to flavor-diagonal soft scalar mass parameters.
Note that the above features hold at the high scale ($\sim$ GUT
scale). RG effects will in general also lead to small flavor
off-diagonal contributions to scalar mass parameters at the
electroweak scale.

The K\"{a}hler potential for moduli fields contains two pieces
\begin{eqnarray} \label{kahler}%
\hat K  = -3\ln (V_{X}) + \frac{2}{V_X}{\rm
Tr}\left(\phi^{\dagger}\phi\right)
\end{eqnarray}
Here $V_X$ is the volume of the $G_2$ manifold in units of the
eleven-dimensional length scale $l_{11}$. The second term
originates from the K\"{a}hler potential for vector-like matter
fields ${\cal Q}$ and ${\cal \tilde{Q}}$ in the hidden sector,
which generally takes the form \cite{Acharya:2008hi}
\begin{equation}
\hat K=\frac{1}{V_X} \left({\cal Q}^{\dagger}{\cal Q}+{\tilde
{\cal Q}}^{\dagger}\tilde {\cal Q}\right)
\end{equation}
By using the D-term equations ${\cal Q}^{\dagger}{\cal Q}={\tilde
{\cal Q}}^{T}\tilde {\cal Q}^{*}$ and the definition of the meson
field $\phi$, it can be rewritten in terms of $\phi$ as given in
second term in Eq.~(\ref{kahler}). Of course, there could be
additional (higher order) corrections, these will be discussed in
Section \ref{general}.
%However, these must be functions of ${\cal Q}^{\dagger}{\cal Q}$
%or ${\tilde {\cal Q}}^{\dagger}\tilde {\cal Q}$ in order to be
%gauge invariant. When written in terms of $\phi$, these
%corrections are always functions of $\phi^{\dagger}\phi$. This
%structure is important for our claim of small CP-violating phases
%as we will see in the rest of the section. Even in the worst case,
%where there might be large corrections to the K\"{a}hler potential
%of the meson fields, the structure of the K\"{a}hler potential
%guarantees that our result is robust.
Now for the simple case $N_f=1$, we can replace ${\rm
det}(\phi^2)$ by $\phi^2$ and ${\rm Tr}(\phi^{\dagger}\phi)$ by
$\overline\phi\phi$ in Eq.~(\ref{superpotential}) and
(\ref{kahler}) respectively. Furthermore $\phi$ can be written as
$\phi=\phi_0\,e^{i\theta}$, with $\theta$ as the phase of $\phi$.

It has been shown in \cite{Acharya:2007rc,Acharya:2008hi} that using the superpotential (\ref{superpotential}) (with $N_f=1$) and
(\ref{kahler}), it is possible to stabilize all moduli, the meson field and one combination of axions. However, at this level the remaining axions are unfixed and remain massless. At the sufficiently long-lived metastable de Sitter minimum of the potential, supersymmetry is spontaneously broken by the strong gauge dynamics and soft supersymmetry breaking terms in the visible sector of the following usual form are generated:
\begin{eqnarray}
&{\cal L}_{soft}& = \frac{1}{2}(M_a \lambda \lambda + h.c.)-
m_{\bar\alpha\beta}^{2}
{\hat C}^{\bar\alpha\dagger}{\hat C}^{\beta} \nonumber\\
&-&\frac{1}{6} {\hat A}_{\alpha\beta\gamma}{\hat C}^{\alpha} {\hat
C}^{\beta}{\hat C}^{\gamma} + \frac{1}{2}\left(B_{\alpha\beta}
{\hat C}^{\alpha}{\hat C}^{\beta} + h.c.\right)
\end{eqnarray}
where ${\hat C}^{\alpha}$s are the canonically normalized chiral
matter fields. The trilinear ${\hat A}_{\alpha\beta\gamma}$ can
often be factorized as
$A_{\alpha\beta\gamma}Y_{\alpha\beta\gamma}$. In the following, we
will be careful in distinguishing between trilinears $\hat A$ and $A$.

\subsection{CP-violating Phases}\label{cpphases}

Now we turn to analyzing the CP-violating phases in the soft
Lagrangian. In order to study the dependence of the soft
parameters on complex phases, it is crucial to understand the
structure of the superpotential in the relevant supersymmetry
breaking vacuum. In the superpotential $W$ in
(\ref{superpotential}), $A_1$, $A_2$, $z_i$ and $\phi$ are complex
variables in general. Without loss of generality, it is possible to choose
$A_1$ and $A_2$ to be real and positive. Then the superpotential (\ref{superpotential})
for $N_f=1$ can be written as : \ba W &=& e^{i\chi_1}\left(|W_1|+|W_2|\,e^{-i(\chi_1-\chi_2)}\right)\\
\chi_{\alpha} &\equiv& b_{\alpha}\sum_{i=1}^N\,N^it_i+\delta_{\alpha1}\,a\,\theta;\;\alpha=1,2\ea
where $|W_1|$ and $|W_2|$ are the magnitudes of the two terms in (\ref{superpotential}).

As explicitly shown in \cite{Acharya:2007rc} and summarized in Appendix \ref{dynamic},
the relative phase between the first and second terms in $W$ is fixed
by the minimization of axions in the vacuum such that: \be \cos{(\chi_1-\chi_2)}=-1\ee
This implies that both terms in the superpotential dynamically align with the same phase (upto a negative sign), leaving just one overall phase in the superpotential, $e^{i\chi_1}$. Since it is possible to do a global phase transformation of the superpotential without
affecting physical observables, this overall phase $\chi_1$ is
not physical and can be rotated away, making the superpotential \emph{real}.
From now on, we will take
$\chi_1=0$. The K\"{a}hler potential, $K$, as seen from
(\ref{kahler}), only depends on real fields $s_i$ which determine
$V_X$ and the combination $\bar{\phi}{\phi}$, so does not contain
any explicit phases.

Note that at this level all but one axions remain massless since the supergravity scalar potential only depends on one linear combination of axions but depends on {\it all} moduli (through the K\"{a}hler derivative). So, one may worry that other possible terms in the superpotential which eventually stabilize the remaining axions will generically stabilize these axions in such a way that the superpotential is not real in the vacuum, thereby ruining the dynamical alignment of phases. However, it turns out that there exists a class of compactifications in which the additional terms in the superpotential stabilizing the remaining axions are exponentially suppressed relative to the first two terms \cite{Acharya:2010zx}. In this class of compactifications, all results of the moduli stabilization mechanism in \cite{Acharya:2007rc, Acharya:2008hi} are kept intact because the higher order terms do not perturb the moduli (and one axion) from their expectation values determined by the first two terms. The remaining axions are stabilized in such a way that the superpotential is real in the vacuum up to exponentially suppressed effects. This is explained in Appendix \ref{dynamic} to which the reader is referred. As shown in \cite{Acharya:2010zx} this mechanism also solves the strong CP problem in an elegant manner, making this mechanism very attractive. Furthermore, it can be shown that the dynamical alignment of phases also works in certain classes of compactifications in Type IIB string theory considered in \cite{Bobkov:2010rf} which have very similar moduli and axion fixing mechanisms.

We now show that with dynamical alignment of phases in the superpotential, there are no CP-violating phases in the soft terms at the high ($\sim$ GUT) scale. The structure of the $F$-terms
$F^I=\hat{K}^{I\bar{J}}F_{\bar{J}}\equiv\hat{K}^{I\bar{J}}(\partial_{\bar
J}{\overline W}+(\partial_{\bar J}K){\overline W})$ where $I,J$
run over both $z_i$ and $\phi$ in general, can be computed as
follows. For $J$ corresponding to M-theory geometric moduli $z_i$,
it is easy to see that $\partial_{\bar J} K$ is real and $\partial_{\bar J}
\overline{W}$ is real (by rotating away the unphysical $\gamma_W$).
For $J$ corresponding to meson moduli $\phi$,
$(\partial_{\bar J}K){\overline W}= {\rm real}\times
e^{i\gamma_{\phi}}$, where $\gamma_{\phi}$ is the phase of $\phi$.
Also, since $W$ depends holomorphically on $\{z_i,\phi\}$ as in
the first line in (\ref{superpotential}), one finds
$\partial_{\bar J} \overline{W}= A_1\,a (\bar{\phi})^{a-1}e^{-b_1 \bar{f}}
=a \overline{W}_1/ \bar{\phi}$,
where $\overline{W}_1$ is the first term in
the complex conjugate of the superpotential (2). Since both terms in the superpotential
have the same phase, again $\partial_{\bar J} \overline{W}={\rm real}\times e^{i\gamma_{\phi}}$.
Therefore, we have $F_{\bar J}={\rm real}$ for $J$ corresponding to the moduli and $F_{\bar J}={\rm real}\times e^{i\gamma_{\phi}}$ for $J$ corresponding to
meson moduli $\phi$.

Now $K^{I\bar{J}}$ is real for $I$ and $J$ both corresponding to either moduli
or meson fields, while for one of them corresponding to moduli and the other corresponding to
the meson field, one has $K^{i\bar{\phi}}={\rm real}\times e^{-i\gamma_{\phi}};
K^{{\phi}\bar{j}}={\rm real}\times e^{i\gamma_{\phi}}$. This can also be verified from the
explicit calculation in \cite{Acharya:2008hi}.
Thus, we see that $F^{I}={\rm real}$ or $F^{I}={\rm
real}\times e^{i\gamma_{\phi}}$ for $I$ corresponding to $z_i$
or $\phi$ respectively. This leads to interesting implications for
the soft supersymmetry breaking parameters.

First, the tree-level gaugino masses are given by:%
\ba \label{gaugino}
M_a^{\mathrm{tree}}(\mu)&=&\frac{g_a^2(\mu)}{8\pi}\,
\left(\sum_{I}e^{\hat{K}/2}F^{I}\partial_{I}\,f_a^{vis}\right)%
\ea %
Since $f_{a}^{vis}$ only depends on the geometric moduli $z_i$
with integer coefficient, and as we have found, the auxiliary
component $F^{I}$ of $z_i$ are real, there are no phases generated
for the tree-level gaugino masses. In the M-theory framework, the
tree-level gaugino masses are suppressed relative to the gravitino
mass \cite{Acharya:2006ia},\cite{Acharya:2008hi}, and the one-loop
anomaly mediated contribution has to be included, which is given
by \cite{Bagger:1999rd}
\begin{eqnarray}
M_{a}^{AMSB} &=& -\frac{g_{a}^{2}}{16\pi ^{2}}\Big( b_a\,
e^{\hat{K}/2}\overline W -b'_a\, e^{{\hat
K}/2}F^{I}\hat{K}_{I}\nonumber\\
&+& 2\sum_{i}C_{a}^{i}e^{{\hat K}/2}F^{I}\partial _{I}\ln
\tilde{K}_{i}\Big)\text{.} \label{anom1}
\end{eqnarray}%
This contribution includes terms proportional to either
${\overline W}$ or $F^{I}\partial_{I} \hat K$ or
$F^{I}\partial_{I} \tilde K_i$. Since the \kahler potential is a
real function of $z_i$, $\partial_{z_i} \hat K$ and
$\partial_{z_i}\tilde K$ are real. In addition, the \kahler
potential only depends on $\bar\phi\phi$, which implies that the
derivative with respective to $\phi$ are proportional to
$\bar\phi\sim e^{-i\gamma_{\phi}}$. Therefore, all these terms are
real, which gives rise to real anomaly mediated gaugino masses.
Hence, the gaugino masses have no observable phase in the above
framework.

The trilinear $A$-terms (with the Yukawa couplings factored
out) are given in general by \cite{Brignole:1997dp}:%
\begin{eqnarray} \label{scalartri}
%A^{\prime }_{\alpha\beta\gamma} &=&
%e^{\hat{K}}\,F^I\,\bigg[\partial_I Y'_{\alpha\beta\gamma} +
%\hat{K}_I Y'_{\alpha\beta\gamma}\nonumber\\
% &-&\left(\tilde{K}^{\delta\bar{\rho}}\,\partial_I
%\tilde{K}_{\bar{\rho}\alpha}\,Y'_{\alpha\beta\gamma}+ (\alpha
%\leftrightarrow \beta)+(\alpha \leftrightarrow \gamma)\right
%)\bigg]\\
A_{\alpha\beta\gamma}&=&e^{\hat
K/2}F^I\partial_I\left[\ln\left(e^{\hat K}
Y'_{\alpha\beta\gamma}/\tilde K_{\alpha}\tilde K_{\beta}\tilde
K_{\gamma}\right)\right]
\end{eqnarray}%
where $I,J$ run over both $z_i$ and $\phi$. It should be noted
that in order to be able to factor out the Yukawa matrices the
matter \kahler metric has to be diagonal. This is a good
approximation in the M-theory framework as we have discussed.
Since the moduli \kahler potential $\hat{K}$ and the visible
sector \kahler metric $\tilde{K}$ are real functions of $z_i+\bar
z_i$ and $\bar\phi\phi$ and superpotential takes the form in Eq.
~(\ref{superW}), it is straightforward to check that the
contractions $F^{I}\partial_{I}\hat{K}$,
$F^{I}\,\partial_{I}\tilde{K}$ and $F^I\partial_I\ln\hat{Y'}$ are
all real, implying that no CP phases are generated in the
trilinear $A$-terms through supersymmetry breaking. However, there
could be phases in the full trilinear couplings $\hat A$ coming
from the Yukawa couplings, as we shall discuss in the next
section.

Finally, we move on to the $\mu$ and $B$ terms. We focus on the
case where the superpotential contribution to the overall high
scale $\mu$ parameter vanishes. This can be easily guaranteed by a
symmetry \cite{Atiyah:2001qf}. In this case, $\mu$ and $B$
parameters of ${\cal O}(m_{3/2})$ can be generated by the
Giudice-Masiero mechanism \cite{Giudice:1988yz} via the parameter
$Z_{\alpha\beta}$ in Eq.~(\ref{kahler}). The general result for
$\mu$ and $B$ can be written in terms of $Z$, $F^{I}\partial_{I}
\hat K$, $F^{I}\,\partial_{I} \tilde K$ and $F^{I}\,\partial_{I}
Z$ \cite{Brignole:1997dp}, all of which have the same phase
$\gamma_Z$ from $Z_{\alpha\beta}$ (complex in general). Therefore,
$\mu$ and $B$ share the same phase $e^{\gamma_Z}$. 
However, this phase is not physical since it can be eliminated by a $U(1)_{PQ}$ rotation \cite{Chung:2003fi}.

Before ending this section, we would like to summarize our result in a more general fashion. 
In the previous analysis, we have seen in general $\gamma_{B} = \gamma_{\mu}$ and $\gamma_{M_a} = \gamma_{A_f}=0$.
%where $\gamma_{B}$ and $\gamma_{\mu}$ are the phases in the $B$ and $\mu$ term respectively, 
Here $\gamma_{M_a}$ and $\gamma_{A_f}$ are defined as the overall phases of $M_a$ and $A_f$ respectively.
Note $\gamma_{A_f}$ is not a basis-independent definition.
From the observational point of view, the relevant physical phases must be reparameterization invariant and basis-independent, 
which can be built from the following combinations \cite{Chung:2003fi}: 
\begin{eqnarray}
\gamma_{1\,\lambda}=\gamma_{\mu}-\gamma_{B}+ \gamma_{M_a}, \quad \gamma_{2\,f}=\gamma_{\mu}-\gamma_{B}+\gamma'_{A_f}\label{phases},
\end{eqnarray}
where $\gamma'_{A_f}=\frac{1}{3}{\rm Arg}[{\rm Det}(\hat A_f Y^{\dagger})]$ is the flavor- and basis-independent CP phase of the trilinears. 
Note that $\gamma'_{A_f} \neq \gamma_{A_f}$ in general unless that the trilinears are universal. 
So we can easily see $\gamma_{1\,\lambda}=0$ and $\gamma_{2\,f}=\gamma'_{A_f}$. This implies that the only flavor-independent phases
that can appear in a physical observable must be those from the trilinears. 
In addition, there can be flavor-dependent CP phases coming from the relative phases of the trilinear matrix elements. 
In the following section, we will discuss the combination of these phases that are relevant for the calculation of EDMs.

\section{CP-violating Phases from Yukawas}\label{textures}

Although the CP-violating phases from supersymmetry breaking are
absent or small as found above, there is an additional
contribution to CP violation if the trilinear $\hat A$ parameters
are not aligned with the Yukawas. This can be easily seen as
follows. Since the Yukawa matrix generically contains ${\cal
O}(1)$ phases in order to explain the observed CKM phase, the
unitary matrices needed to go to the super-CKM basis (in which the
Yukawa matrices are real and diagonal) also contain some phases.
Therefore, the rotation by itself can induce CP-violating phases
even if the $A$ or $\hat A$ matrices are initially completely real
as long as $\hat A$'s are not proportional to Yukawas in the
flavor eigenstate basis (or equivalently $A$'s are flavor
non-universal and non-diagonal). This implies in particular that
the diagonal components of trilinear $\hat A$ will contain CP
phases in the super-CKM basis, giving rise to possibly important
contributions for EDMs.

In the M-theory framework, the Yukawa couplings $Y'_{\alpha\beta\gamma}$
depend holomorphically on the geometric moduli
$z_i$ in general which get non-zero $F$-term vevs. Hence, from
(\ref{scalartri}) we find that the second term in the expression
for trilinears gives rise to an ${\cal O}(1)$ misalignment between
the Yukawas and the trilinears. If the Yukawa couplings depend on
moduli or other hidden sector fields which do not break
supersymmetry, then the trilinears can be naturally aligned with
the Yukawas \cite{Conlon:2007dw}. However, within M-theory, this
does not seem to be a generic situation; hence we will consider
the conservative case in which the trilinears are misaligned with
the Yukawas.

In the remainder of this section, we will estimate the diagonal CP
phases in the trilinear $\hat A$ in the super-CKM basis since they
are directly related to the EDM observables. We consider flavor
non-universal and non-diagonal trilinear $A$-matrices (in the
flavor eigenstate basis) at the GUT scale with real ${\cal O}(1)$
matrix elements. To set the conventions, we write down the soft
trilinear terms explicitly
\begin{eqnarray}
{\cal L}_{soft}& \sim & A_{ij}^{u}\,Y_{ij}^{u}\,\bar Q_{Li}H_u
u_{Rj} +
A_{ij}^{d}\,Y_{ij}^{d}\,\bar Q_{Li}H_d d_{Rj} \nonumber\\
&+& A_{ij}^{e}\,Y_{ij}^{e}\, \bar L_{Li}H_d e_{Rj}
\end{eqnarray}
where $A^{u,d,e}$ are the trilinear matrices in the gauge
eigenstate basis of matter fields.

In the M-theory framework, chiral matter fields are localized on
singular points inside the compact $G_2$ manifold
\cite{Atiyah:2001qf,Witten:2001uq,Acharya:2001gy,Berglund:2002hw}.
Although a detailed understanding of Yukawas within M-theory is not yet available,
a hierarchical Yukawa texture seems well motivated. From a phenomenological point of
view, therefore, we consider the following Yukawa texture:
\begin{equation}%
Y_{ij}^{u}\sim \epsilon_i^q \epsilon_j^u, \quad Y_{ij}^d\sim
\epsilon_i^q \epsilon_j^d, \quad Y_{ij}^{e} \sim \epsilon_i^l
\epsilon_j^e, \label{factorization}
\end{equation}%
which can arise naturally from the localization of
matter fields in extra dimensional models
\cite{ArkaniHamed:1999dc,Mirabelli:1999ks,Kaplan:2001ga,Abe:2004tq}
or from a spontaneously broken flavor symmetry (Froggatt-Nielson
mechanism) \cite{Froggatt:1978nt}. Then, the fermion mass
hierarchy is given by:
\ba %
&& m_i^u/m_j^u \sim |\epsilon_i^q \epsilon_i^u|/|\epsilon_j^q
\epsilon_j^u|, \quad m_i^d/m_j^d \sim |\epsilon_i^q
\epsilon_i^d|/|\epsilon_j^q \epsilon_j^d|, \nonumber\\
&&m_i^e/m_j^e \sim |\epsilon_i^l \epsilon_i^e|/|\epsilon_j^l
\epsilon_j^e|
\ea%
It is straightforward to check that the observed fermion mass
hierarchy can be accommodated by a set of properly chosen
$\epsilon_i$ with the hierarchy $|\epsilon_1| \lesssim
|\epsilon_2| \lesssim |\epsilon_3|$. The above Yukawa couplings
can have ${\cal O}(1)$ phases in order to explain the CP phase in
the CKM matrix. To simplify the discussion, we eliminate the
phases in the diagonal elements by a redefinition of the quark and
lepton fields. Therefore, the diagonal elements $(\hat
A^{\psi})_{11,22,33}$ with $\psi=u,d,e$ are all real at the GUT
scale.

First, we point out that the renormalization group (RG)
corrections to the trilinear couplings typically mix the phases
between different flavors. This will lead to phases in the
diagonal elements of the trilinear matrices. It can be understood
from the RG equation for Yukawa couplings and trilinear couplings,
e.g. for $Y^u$ and ${\hat A}^u$, which are given by
\begin{eqnarray}
\frac{d Y^u}{dt} &\sim & \frac{1}{16\pi^2} Y^u \left[ 3 {\rm
Tr}(Y^u Y^{u\dagger}) + 3 Y^{u\dagger}Y^u + Y^{d\dagger} Y^d
%\nonumber\\ &-& \frac{16}{3}g_3^2 -3 g_2^2 - \frac{13}{15} g_1^2
\right]\nonumber\\
\frac{d {\hat A}^u}{dt} &\sim & \frac{1}{16\pi^2} {\hat A}^u
\left[ 3 {\rm Tr}(Y^u Y^{u\dagger}) + 5 Y^{u\dagger}Y^u +
Y^{d\dagger} Y^d
%\nonumber\\ &-& \frac{16}{3}g_3^2 -3 g_2^2- \frac{13}{15} g_1^2
\right]\nonumber\\
&+& \frac{Y^u}{16\pi^2} \left[ 6 {\rm Tr}({\hat A}^u
Y^{u\dagger})+ 4Y^{u\dagger}{\hat A}^u +
2 Y^{d\dagger}{\hat A}^d %\nonumber\\
%&+& \frac{32}{3}g_3^2 M_3 + 6 g_2^2 M_2 + \frac{26}{15}g_1^2 M_1
\right]
\end{eqnarray}
where only terms involving Yukawas are explicitly shown. From the
above equations, we notice that the phases in ${\hat A}^u$ evolve
during the RG running. To illustrate this, one can examine the
following term which contributes to the running of ${\hat
A}^u_{11}$:
\begin{eqnarray}
\frac{d {\hat A}^u_{11}}{dt}\sim \frac{5}{16\pi^2} {\hat A}^u_{13}
Y^{u\dagger}_{33}Y^u_{31}+
\frac{4}{16\pi^2}Y^u_{13}Y^{u\dagger}_{33}{\hat A}^u_{31}
\end{eqnarray}
From the equation, one can see that the phases of $Y^u_{31}$,
$Y^u_{13}$, ${\hat A}^u_{13}$ and ${\hat A}^u_{31}$ can enter that
of ${\hat A}^u_{11}$ through RG effects although it was real at
the high scale. The magnitude of this correction can be
significant since the magnitude of the right-hand side of the
above equation is proportional to $\frac{1}{16\pi^2} |Y^u_{33}|^2$
given the factorizable Yukawa matrices as in
Eq.~(\ref{factorization}). This indicates that the RG corrections
to ${\hat A}^u_{11}$ can give ${\cal O}(1)$ phases. This is also
true for other elements in the trilinear matrices and Yukawa
matrices. The only exception is for the third generation
$Y^u_{33}$ and ${\hat A}^u_{33}$, for which the largest RG
corrections come from the terms involving only $Y^u_{33}$ and
${\hat A}^u_{33}$ with additional flavor mixing terms typically
suppressed by $\epsilon_{2}^2/\epsilon_{3}^2$. Since ${\hat
A}^u_{33}$ has the same phase as $Y^u_{33}$, the corresponding
A-term $A^u_{33}={\hat A}^u_{33}/Y^u_{33}$ remains real up to
corrections of the order $(\epsilon_{2}/\epsilon_{3})^2$.

Starting from the Yukawa matrices $Y_{ij}^{u,d,e}$ defined in the
flavor eigenstate basis, the super-CKM basis can be achieved by
unitary rotations of the matter fields so that the Yukawa matrices
are real and diagonal. In the super-CKM basis, the trilinear
couplings become
\begin{eqnarray}%
({\hat A}^{\psi}_{\rm SCKM})_{ij}= (V_L^{\psi\dagger})_{il}\,
A^{\psi}_{lk}\,Y^{\psi}_{lk}\,(V_R^{\psi})_{kj}\label{normalizedA}
\end{eqnarray}%
where $\psi= u,d,e$. To be concrete, we focus on the up-type trilinear
in the discussion below. 
Given the hierarchical Yukawa matrices in
Eq.~(\ref{factorization}), the unitary transformation matrices are
given by
\ba%
(V_L^{u})_{ij}&\sim& (V_L^{u})_{ji}\sim \epsilon_{i}^{q}
/\epsilon_{j}^{q}, \; {\rm for} \; i < j \nonumber\\
(V_R^{u})_{ij}&\sim& (V_R^{u})_{ji}\sim
\epsilon_{i}^{u}/\epsilon_{j}^{u}, \; {\rm for} \; i < j
\ea%
One can now perform the same transformation for the trilinear
terms to get the diagonal elements in the super-CKM basis, which
can be schematically written as
\begin{eqnarray}%
({\hat A}^{u}_{\rm SCKM})_{11}&\approx & \epsilon_1^q\epsilon_1^{u}\sum_{i,j=1,2,3}\xi_{ij}  A_{ij}^{u}, \label{A1}\\
({\hat A}^{u}_{\rm SCKM})_{22}&\approx & \epsilon_2^q\epsilon_2^{u} \sum_{i,j=2,3}\eta_{ij} A_{ij}^{u}, \label{A2}\\
({\hat A}^{u}_{\rm SCKM})_{33}&\approx & Y^u_{33} A^{u}_{33}\sim
\epsilon_3^q\epsilon_3^u A_{33}^{u} \label{A3}
\end{eqnarray}%
where $\xi_{ij}$ and $\eta_{ij}$ are possibly ${\cal O}(1)$
coefficients arising from those implicit coefficients in the Yukawa matrices in (\ref{factorization}), and are complex in
general. In the above equations, we have neglected subleading
terms suppressed by $\epsilon_i/\epsilon_j$ for $i < j$. Since the
off-diagonal components in $A_{ij}$ can be ${\cal O}(1)$ within our
framework, the summations in Eq.~(\ref{A1}) and (\ref{A2})
can be of ${\cal O}(1)$ in magnitude with ${\cal O}(1)$
phases. The $(\hat A)_{33}$ component, however, does not mix with
other components and is proportional to $Y^u_{33}$, so no phase is
generated at leading order for ${\hat A}^{u}_{33}$.

Therefore, we conclude that the first two diagonal components of
the complete trilinear coupling in the super-CKM basis can contain
order one phases, while the third diagonal component is real up to
small corrections, i.e.
\begin{eqnarray}\label{suppression}
{\rm Im}({\hat A}^{\psi}_{\rm SCKM})_{11} &\sim& A_0 Y^{\psi}_{11}, \nonumber\\
{\rm Im}({\hat A}^{\psi}_{\rm SCKM})_{22} &\sim& A_0 Y^{\psi}_{22}, \nonumber\\
{\rm Im}({\hat A}^{\psi}_{\rm SCKM})_{33} &\sim& A_0
\left(\frac{\epsilon_2}{\epsilon_3}\right)^2 \, Y^{\psi}_{33}
\end{eqnarray}
where $A_0$ is the characteristic magnitude of the trilinear
A-terms. Here $(\epsilon_2/\epsilon_3)^2$ is roughly the ratio of 
the second and third generation fermion masses, which depends on the choice of $\psi=u,d,e$. 
In the discussion of EDMs in Subsection \ref{sec:two-loop}, we will be taking $\epsilon_2/\epsilon_3 \sim 0.1$ 
as an order of magnitude estimate compatible with the quark mass hierarchy. 

Although the result in Eq.(\ref{suppression}) is derived for the specific class of Yukawa matrices in 
Eq.(\ref{factorization}), it is nevertheless more generic than that. In fact, for any Yukawa
matrices such that the linear combination of terms giving the diagonal component of $Y_{\rm SCKM}$ do not have any large cancellation,
the first two relations in Eq.(\ref{suppression}) would still be valid. An additional requirement that the flavor mixing contribution to 
$(Y_{\rm SCKM})_{33}$ is suppressed would lead to a similar relation as the last one in Eq.(\ref{suppression}). 
These conditions can be easily accommodated for more general Yukawa textures. 
Other Yukawa textures that do not satisfy these conditions are possible, 
but seem less generic and natural in order to get the hierarchical fermion masses and mixings. 
This result has important implications for EDM predictions, which we shall discuss in section \ref{predict}.

\section{Electric Dipole Moments and The Experimental
Limits}\label{edm}

Before starting our calculation of EDMs, we briefly summarize some
general results relevant for the calculation of EDMs. In the
minimal supersymmetric standard models, the important CP-odd terms
in the Lagrangian are:
\begin{eqnarray}
\delta{\cal L}=&& -\sum_{q=u,d,s} m_q \bar q(1+i\theta_q\gamma_5)q
+ \theta_G \frac{\alpha_s}{8\pi}G\tilde G
\nonumber \\
&&-\frac{i}{2}\sum_{f=u,d,s}(d_{q}^{E} \bar q
F_{\mu\nu}\sigma_{\mu\nu}\gamma_5 q +
\tilde d_{q}^{C} \bar q g_s t^aG_{\mu\nu}^a\sigma_{\mu\nu} \gamma_5 q)\nonumber\\
&& - \frac{1}{6}d_q^G
f_{\alpha\beta\gamma}G_{\alpha\mu\rho}G_{\beta\nu}^{\rho}G_{\gamma\lambda\sigma}
\epsilon^{\mu\nu\lambda\sigma}, \label{CPV}
\end{eqnarray}
where $\theta_G$ is the QCD $\theta$ angle,  The terms in the
second line in (\ref{CPV}) are dimension five operators, which are
generated by CP violation in the supersymmetry breaking sector and
evolved down to $\sim 1$ GeV. The coefficients $d_{q}^{E,C}$
correspond to quark electric dipole moment and chromo-electric
dipole moment(CEDM) respectively. The last line in (\ref{CPV})
contains the gluonic dimension six Weinberg operator. The CP-odd
four-fermion interactions are not important here, and so have not
been included above.

Now let us briefly summarize the EDM results for electrons,
neutrons and mercury in terms of the coefficients of these
operators. The electron EDM in minimal supersymmetric models is
given by:
\begin{equation}
d_e^E = d^{\chi^+}_e + d_e^{\chi^0}+ d_e^{\rm BZ}\nonumber
\end{equation}
where $d_e^{\chi^{\pm}}$ and $d_e^{\chi^0}$ are one-loop
contributions from the neutralino and chargino while $d_e^{\rm
BZ}$ is the two-loop Barr-Zee type contribution
\cite{Barr:1990vd,BowserChao:1997bb,Chang:1998uc,
Pilaftsis:1999td,Chang:1999zw,Chang:2002ex,Pilaftsis:2002fe,Chang:2005ac,Li:2008kz}.
It should be noted that what is actually measured is the atomic
EDM $d_{Tl}$, which receives contributions mainly from the
electron EDM and the CP-odd electron-nucleon couplings
\cite{Pospelov:2005pr}: %
\ba %
d_{Tl}&=& -585\times d_e^E - 8.5\times 10^{-19} e\,{\rm
cm}(C_S\,{\rm TeV}^2) + \cdots\nonumber
\ea%
where $C_S$ is the coefficient of the operator $\bar e i\gamma_5
e\bar N N$. The $C_S$ coefficient could be generated from a new
scalar particle coupled to quarks and leptons through a CP-odd
higgs like coupling \cite{Pospelov:2005pr}. However, this is
independent of CP-odd interactions originating from the soft
terms. Given the current experimental limit
$|d_{Tl}|<9\times10^{-25} e\,\rm{cm}$, we obtain an upper limit on
electron EDM
\begin{equation}
|d_{e}^E| < 2 \times10^{-27} e\,\rm{cm}\nonumber
\end{equation}

For the neutron, there exist several different approaches to
compute the corresponding EDM. In the following discussion, we
shall follow a simple approach, i.e., the naive dimensional
analysis
(NDA)\cite{Manohar:1983md,Arnowitt:1990eh,Arnowitt:1990je}.
The neutron EDM can be calculated as:%
\begin{equation}\label{neutron}
d_n =\frac{4}{3} d_d -\frac{1}{3} d_u.
\end{equation}%
In this expression, the quark EDMs can be estimated via NDA as:%
\begin{equation}
d_q = \eta^E d_q^E + \eta^C \frac{e}{4\pi} d_q^C + \eta^G
\frac{e\Lambda}{4\pi} d^G\nonumber
\end{equation}%
with $d_q^{E,C} = d_q^{\tilde g (E,C)} + d_q^{\tilde \chi^{+}
(E,C)} + d_q^{\tilde\chi^0 (E,C)}$. The QCD correction factors are
given by $\eta^E=1.53$, $\eta^C \sim \eta^G \sim 3.4$
\cite{Ibrahim:1997gj}, and $\Lambda\sim 1.19\GeV$ is the chiral
symmetry breaking scale. The current experimental limit on neutron
EDM is given by
\begin{equation}
|d_{n}| < 3\times 10^{-26} e\,\rm{cm}\nonumber
\end{equation}

The current theoretical estimate for the mercury EDM induced by
dimension $5$ operators is given by \cite{Demir:2003js}:
\begin{equation}\label{HgEDM}
d_{H_g} = - 7.0\times 10^{-3}\, e \,(d_d^C - d_u^C- 0.012 d_s^C) +
10^{-2}\times d_e\nonumber
\end{equation}
where we have included the contribution from the strange quark
CEDM \cite{Falk:1999tm}. The recent experimental result on Mercury
EDM \cite{Griffith:2009zz} significantly tightens the bound
\begin{equation}
|d_{Hg}|  < 3.1 \times 10^{-29}\,e\,\rm{cm}\nonumber
\end{equation}

In the Standard Model, the primary source of hadronic EDMs comes
from the QCD $\theta$-term in (\ref{CPV}). This gives the following
results~\cite{Baluni:1978rf,Crewther:1979pi,Pospelov:1999ha,Pospelov:2005pr}:
\begin{eqnarray}
d_n &\sim& 3 \times 10^{-16} \theta \;\;\,e\,\rm{cm}\nonumber\\
d_D &\sim& -1 \times 10^{-16} \theta \;\;\,e\,\rm{cm}\nonumber\\
|d_{Hg}| &\sim& {\cal O}(10^{-18}- 10^{-19})\,\theta \;\;
\,e\,\rm{cm}\label{sm-edm}
\end{eqnarray}
On the other hand, the electron EDM is induced by the SM electroweak
interactions, which is typically of order $10^{-38}\,e\,\rm{cm}$~
\cite{Archambault:2004td,Pospelov:1991zt}.
The results in~(\ref{sm-edm}) together with the suppressed leptonic EDMs provide a
correlation pattern for the $\theta$-induced electric dipole
moments. The current upper bound on the neutron EDM implies
$\theta < {\cal O}(10^{-10})$, which leads to the strong CP
problem. Once EDMs are observed for n, Hg and Tl it will be
essential to separate the strong and weak contributions, by
combining data on different nuclei and $d_e^E$.

\section{Predictions for EDMs}\label{predict}

For an explicit computation of the EDMs, it is important to
specify the general structure of supersymmetry breaking
parameters, in particular the structure of the trilinear
parameters (especially the imaginary part of the diagonal
components), as well as that of the scalar and gaugino masses,
since all of these appear in the final expression for the EDMs.
This is the subject of this section.

Within the M-theory framework, the general structure of
supersymmetry breaking parameters is as follows. For the choice of
microscopic parameters with a vanishingly small positive
cosmological constant, the gravitino mass naturally turns out to
be in the range $10$-$100$~TeV \cite{Acharya:2007rc}. The
gravitino mass is essentially $\sim F_{\phi}/m_p$. However, as
mentioned earlier, the $F$-terms of the moduli are suppressed
compared to $F_{\phi}$. Since the gauge kinetic function for the
visible sector depends only on the moduli, from (\ref{gaugino}) it
is easy to check that the gaugino masses are suppressed relative
to that of the gravitino. However, this suppression does not hold
for the scalar masses, trilinears, $\mu$ and $B\mu$ parameters
unless the visible sector is sequestered from the supersymmetry
breaking sector. Sequestering does not seem to be generic in M-theory \cite{Acharya:2008hi},
so scalar masses, trilinears, $\mu$ and $B\mu$ parameters
typically turn out to be of ${\cal O}(m_{3/2}) \sim {\cal
O}(10)$~TeV. The third generation squarks, however, could be
significantly lighter because of the RG effects.

As we have discussed in Section \ref{leading}, within the M-theory
framework it is natural to expect that the K\"{a}hler metric for
visible matter fields is approximately diagonal in the flavor
indices. Then, the scalar mass matrix turns out to be roughly
diagonal with suppressed off-diagonal contributions. The estimates
for the EDMs then depend on the overall scale of the squark
masses. So, for concreteness we consider gauginos with masses
$\lesssim$ 600 GeV, non-universal but flavor-diagonal scalar mass
matrices with masses $\sim 20$ TeV, and $\mu, B\mu$ and trilinear
parameters of the same order as scalar masses. Some contributions
to EDMs depend primarily on third generation sfermion masses, so
we also mention the situation when third generation scalars are
much lighter, i.e. ${\cal O}(1)$ TeV.

We now estimate the contribution to the EDMs of the electron,
neutron and mercury from dimension $5$ and $6$ operators
(Eq.~(\ref{CPV})) in the M-theory framework. As we have seen in
the Section III, the CP-violating phases appear only in the
trilinear $\hat{A}$ parameters. After renormalization group
evolution and the super-CKM rotation of the trilinear matrices,
these phases appear in the off-diagonal elements in the squark
mass matrices, leading to
imaginary parts of the following mass-insertion parameters: %
\ba (\delta_{q}^{ii})_{LR}
&=&\frac{v_{q}((\hat{A}^{q}_{SCKM})_{ii}-\mu^{*}
Y^{q}_{ii}R_{q})}{(m_{\tilde{q}}^2)_{ii}} %
\ea %
where $R_{u(d)}=\cot\beta\;$($\tan\beta)$ and
$v_{u(d)}=v\sin\beta\;(v\cos\beta)$. As explained above,
$\hat{A}_{SCKM}$ is in general a $3\times3$ matrix in the
Super-CKM basis and its diagonal components contain CP-violating
phases. Thus, these insertion parameters contribute to EDMs
through the dimension $5$ and $6$ operators in (\ref{CPV}).

\subsection{Leading Contributions}

\begin{figure}
\begin{centering}
\includegraphics[width=2in]{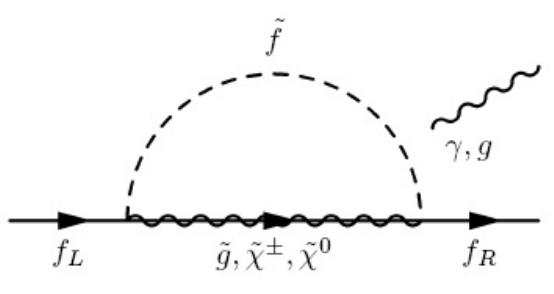}
\par\end{centering}
\caption{One-loop contributions to fermion (C)EDMs.}
\centering{}\label{1loop}
\end{figure}

The dimension five electric and chromo-electric couplings can be
generated at leading order
\cite{Masiero:1997bn,Ibrahim:1997nc,Ibrahim:1997gj} at one-loop
through the vertices $f\tilde{f}\tilde{\chi}_{i}^{0}$,
$f\tilde{f}'\tilde{\chi}_{i}^{\pm}$ and $q\tilde{q}\tilde{g}$ as
can be seen in Fig. \ref{1loop}.

\begin{figure}
\begin{centering}
\includegraphics[width=8.cm,height=5.5cm]{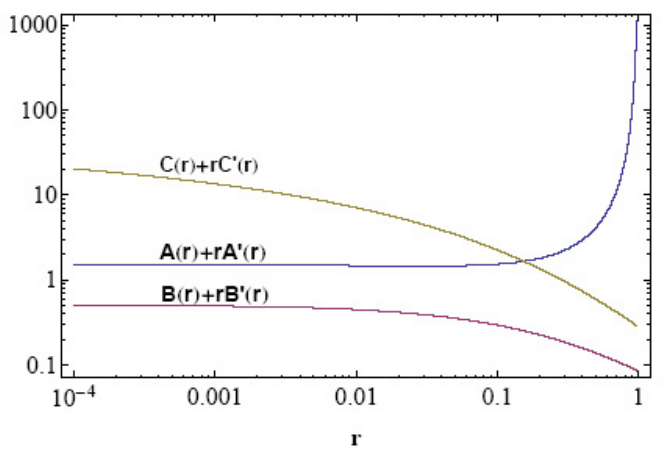}
\par\end{centering}
\caption{Comparison of the one-loop functions $A(r)$, $B(r)$ and
$C(r)$. The $x$ coordinate is the ratio $r\equiv
m_{i}^{2}/m_{\tilde{q}}^{2}$.} \centering{}\label{loop-function}
\end{figure}

First consider the quark CEDM which contributes to both the
mercury and neutron EDMs. Since there exists a hierarchy between
gauginos and squarks in the M-theory
framework\cite{Acharya:2006ia,Acharya:2007rc}, one can expand
using the small ratio $r\equiv m_{i}^{2}/m_{\tilde{q}}^{2}$, where
$m_{i}$ is the corresponding neutralino, chargino or gluino masses
in the diagram. One then obtains the following result%
\begin{equation}\label{CEDM}
d_{q}^{C}\sim\frac{g_{s}\alpha_s}{4\pi}\frac{m_{q}}{m_{i}^{3}}{\rm
Im}({A}^{q}_{\rm SCKM})r^{2}G(r)
\end{equation}%
where $A^q_{SCKM}$ is the diagonal element of the corresponding
trilinear matrix (factoring out the Yukawa coupling) in the
super-CKM basis. In the expression, the function
$G(r)=C(r)+rC'(r)$ for gluinos and $G(r)=B(r)+rB'(r)$ for
charginos and neutralinos. The function $B(r)$ and $C(r)$ are loop
functions defined in the Appendix \ref{leading}. One can see that
$d_{q}^{C}$ decreases rapidly as $m_{\tilde{q}}^{-4}$ when the
squark masses increase. However, the function $G(r)$ behaves
differently for different particles
($\tilde{g},\tilde{\chi}^{\pm},\tilde{\chi}^0$) in the loop. Due
the gaugino and squark mass hierachy, $r$ is small. From Fig.
\ref{loop-function}, we can see that $C(r)+rC'(r)$ is enhanced in
the small $r$ region compared to other functions which remain
small. Therefore, the gluino contribution dominates the quark
CEDM. For the quark EDM, it is given by a similar expression as
(\ref{CEDM}) but now the quantity $G(r)$ is determined only by
$A(r)$ and $B(r)$. In particular, $G(r)$ is determined solely by
$B(r)$ for $\tilde{g}$ and $\tilde{\chi}^0$ in the loop, and by a
combination of $A(r)$ and $B(r)$ for $\tilde{\chi}^{\pm}$ in the
loop. Since $A(r)+rA'(r)$ and $B(r)+rB'(r)$ are much smaller than
$C(r)+rC'(r)$ as seen from Figure \ref{loop-function}, the quark
EDM contributions to the neutron EDM are negligible compared to
that of the quark CEDM contributions. Therefore, we only need to
calculate the quark CEDM, for which the gluino diagram gives the
dominant contribution as explained above. Since $A^{q}_{SCKM}\sim
m_{\tilde{q}}$ in the M-theory framework, one obtains:
\begin{equation} %
d_{q}^{C} \sim 10^{-28}\cdot \left(\frac{m_{q}}{1{\rm
MeV}}\right)\left(\frac{m_{\tilde g}}{600{\rm
GeV}}\right)\left(\frac{20{\rm
TeV}}{m_{\tilde{u}}}\right)^{3} e\,\rm{cm}%
\end{equation}

Based on the quark EDM and CEDM, the neutron EDM can be computed from (\ref{neutron}):
\begin{eqnarray}
d_{n}^{NDA}\sim  3\times10^{-28} \cdot \left(\frac{m_{\tilde
g}}{600{\rm GeV}}\right)\left(\frac{20{\rm
TeV}}{m_{\tilde{u}}}\right)^{3}\,e\,\rm{cm}
\end{eqnarray}
Similarly, the mercury EDM is  %
\begin{eqnarray}
|d_{H_{g}}|\sim 10^{-30}\cdot \left(\frac{m_{\tilde g}}{600{\rm
GeV}}\right)\left(\frac{20{\rm TeV}}{m_{\tilde{u}}}\right)^{3}
\,e\,\rm{cm}
\end{eqnarray}%

Moving on to the electron EDM, it can be computed at leading order
from the one-loop neutralino and chargino diagrams. First, we
notice that the chargino diagram does not contribute in our framework. This is because
the fact that the $\mu$-term phase\footnote{Here we have set $\gamma_{B}=0$ by a $U(1)_{PQ}$ rotation.} and the gaugino phases are zero and there is no phase
in the chargino mixing matrix as can be seen from Eq.(\ref{chargino}) in the Appendix.
%due to zero phases in the $\mu$-term and the gaugino masses $M_a$ and therefore in the chargino mixing matrices 
The neutralino contribution, on the other hand, gives rise to a non-zero
contribution because of a dependence on the selectron mixing
parameters (which contains CP-violating phases) in its couplings
(see Eq. (\ref{neutralino})). Given the fact that the higgsino coupling to
electron and selectron is suppressed by the small electron Yukawa coupling,
and the wino does not couple to right-handed fermions and sfermions, the
dominant contribution is from the diagram with $\tilde
\chi_2^0$(almost pure bino in the M-theory framework), which can
be calculated using Eq.~(\ref{de-neutralino}) in the Appendix.
Thus, the electron EDM is given by:
\begin{equation}%
d_{e}^{E}\sim\left(\frac{m_{{\tilde \chi}_2^0}}{200{\rm
GeV}}\right)\left(\frac{20{\rm
TeV}}{m_{\tilde{e}}}\right)^{3}\times10^{-31}\,e\,\rm{cm}
\end{equation}

\subsection{Two-loop Contributions}\label{sec:two-loop}

\begin{figure}
\begin{center}
\includegraphics[width=2.5in]{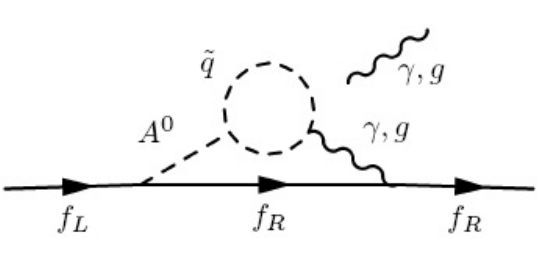}
\par\end{center}
\caption{Two-loop Barr-Zee type diagrams contributing to fermion
(C)EDMs.} \label{2loop}
\end{figure}

\begin{figure}
\begin{center}
\includegraphics[width=2.5in]{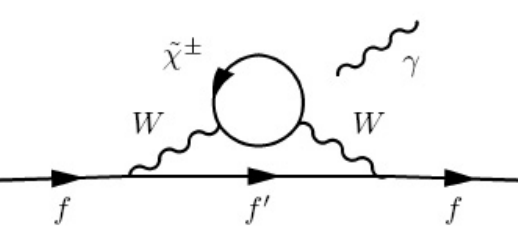}
\par\end{center}
\caption{Two-loop Barr-Zee type diagrams which do not involve
sfermion in the loop.} \label{2loop-ww}
\end{figure}

So far, we have considered the one-loop contribution to quark and
electron EDMs (and/or CEDMs). In addition, there are two-loop
Barr-Zee type
contributions~\cite{Barr:1990vd,BowserChao:1997bb,Chang:1998uc,
Pilaftsis:1999td,Chang:1999zw,Chang:2002ex,Pilaftsis:2002fe,Chang:2005ac,Li:2008kz}
such as the one in Fig.~\ref{2loop}. In general, the Barr-Zee type
diagrams can involve squarks, charginos or neutralinos in the
inner loop, and higgs bosons (neutral or charged) and/or gauge
bosons in the outer loop (the two-loop diagram considered
in split supersymmetry is not relevant here, since there the CP
violation is not from trilinear couplings, but instead from the
chargino sector). Since only the trilinear couplings contain
CP-violating phases in our framework, we consider those diagrams
with squarks running in the inner loop as seen in Figure
\ref{2loop}. One might wonder whether there are any two-loop
diagrams that would contribute if there were phases in the gaugino
masses or $\mu$ term such as in the split
supersymmetry scenario\cite{Chang:2005ac,Giudice:2005rz}. Since
the higgsino in the M-theory framework is very heavy with mass
$\mu\sim m_{3/2}$ and hence decoupled from the low energy theory,
the only diagram which might contribute is the one in
Fig.~\ref{2loop-ww}. However, it turns out that the CP phases in
the two W-chargino-neutralino couplings cancel out (up to a small
correction due to the heavy higgsino) in the final result giving
no EDM contribution.

%We should emphasize that it may be important to include other generation squarks in
%the calculation in certain cases.
%These contributions are important for large CP-violating phases in
%the third generation squarks since the third generation squarks
%couple dominantly to the higgs pseudoscalar ($A_0$) in the loop
%due to large yukawa couplings.
When the mass splitting between the two third generation squarks
is not particularly large, the diagram to the quark CEDM can be
estimated as (see Appendix \ref{appendix:barr-zee}):
\begin{eqnarray}
d_{f}^{C\,{\rm BZ}} &\approx& \frac{g_{s}\alpha_{s}}{64\pi^{3}}
\frac{m_f R_f \mu}{M_{A}^{4}}\sum_{q=\tilde t,\tilde b} y_{q}^{2}\,{\rm Im}(A^{q}_{\rm SCKM})F'(r_q)\nonumber\\
&\sim&  10^{-32}\cdot R_f\left(\frac{m_{f}}{1{\rm
MeV}}\right)\left(\frac{20{\rm TeV}}{m_{\tilde{q}}}
 \right)^{2} \,e\,{\rm cm}\label{Eq:Barr-Zee1}
\end{eqnarray}
where $R_f=\cot\beta$($\tan\beta$) for $I_3=1/2$($-1/2$),
$r_q\equiv m_{\tilde q}^2/M_A^2$ with $m_{\tilde{q}}$ third
generation squark mass and $M_A$ the pseudoscalar mass of $A_0$.
Here we have used (\ref{suppression}) for ${\rm Im(A^q_{SCKM})}$. For simplicity,
we also take $\mu \sim M_{A}
\sim m_{\tilde t,\tilde b} \sim m_{\tilde u}$. It can be seen that
the result of the Barr-Zee diagram to quark CEDM (similar for EDM)
is negligibly small. One of the reasons is that CP violation in
the third generation is suppressed by about two orders of
magnitude as in (\ref{suppression}).
Similarly, for the electron EDM the result is:%
\begin{equation}
d_{e}^{E\,{\rm BZ}}\sim 10^{-33}\cdot\left(\frac{20{\rm
TeV}}{m_{\tilde{u}_3}}\right)^{2}\tan\beta\;\;e\,\rm{cm}\label{Eq:Barr-Zee2}
\end{equation}%
which is again quite suppressed. %However, when the third
%generation scalar masses are much lighter ($\sim 1$ TeV) than the
%first two generations, such as from RG effects, the Barr-Zee
%diagram could give rise to large contributions.
This contribution may be enhanced for large $\tan \beta$ as seen
from above. The M-theory framework, however, generically predicts $\tan
\beta={\cal O}(1)$\cite{Acharya:2008hi}.

The neutron EDM could also get a contribution from the dimension
six pure gluonic operator (Weinberg operator), which can be
generated from the two loop gluino-top-stop and
gluino-bottom-sbottom diagrams. For the case where CP violation
only comes from the soft trilinear couplings, the result can be
estimated by \cite{Dai:1990xh}
\begin{equation}
d^{G}\approx -3\alpha_{s}\left(\frac{g_{s}}{4\pi}\right)^{3}
\frac{1}{m_{\tilde{g}}^{3}}\sum_{q=t,b}{\rm Im}(A_{q}^{\rm
SCKM})z_{q}H(z_{1},z_{2},z_{q})
\end{equation}
where $z_{i}=m_{{\tilde{q}}_{i}}^2/m_{\tilde{g}}^{2}$ for $i=1,2$,
and $z_{q}=m_{q}^2/m_{\tilde{g}}^{2}$ for $q=t,b$. The two-loop
function $H(z_{1},z_{2},z_{t})$ is given in \cite{Dai:1990xh}.
This gives a contribution to the neutron EDM $d_{n}^{G}\sim
10^{-30}\,e\,{\rm cm}$ for $m_{\tilde{t},\tilde{b}}\approx
20$~TeV, $m_{\tilde{g}}=600$~GeV and $A_{q}= 20$~TeV. Thus the
neutron EDM from the Weinberg operator is smaller than the
one-loop CEDM contribution. However, when the masses of the third
generation squarks and trilinears are around $1$~TeV, the
contribution to the neutron EDM can be significantly larger, and
be comparable to the one-loop result.

\begin{figure}
\begin{center}
\includegraphics[width=1.8in]{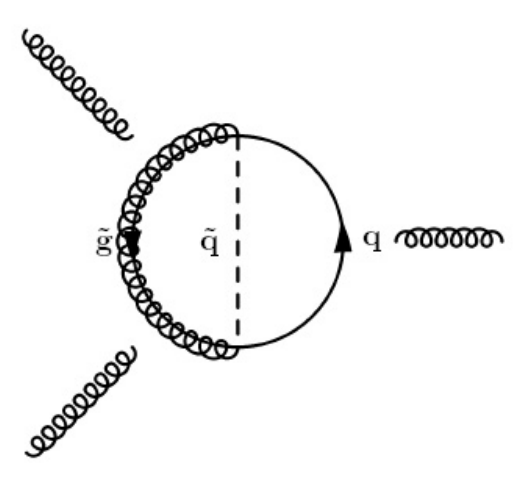}
\par\end{center}
\caption{Two-loop diagrams contributing to the Weinberg operator.}
\label{Weinberg}
\end{figure}

%In general, imaginary parts of the off-diagonal components of
%squark mass matrix, e.g. non-zero ${\rm Im}(m_{{\tilde q}\,
%23}^{2})_{LR}$ could also generate quark and electron EDMs.
%However, these contributions require more than a single insertion,
%therefore they are usually not as large as the one-loop
%contribution we have discussed.

To summarize our results, we have calculated the EDMs arising from
the CP-violating phases in the trilinear terms in a general
framework with light gauginos and heavy sfermions, and the results
are within current experimental bounds. We find that the one-loop
diagram is typically the dominant contribution to EDMs. However,
in contrast to the situation in which gaugino and sfermion masses
are comparable, the one-loop diagram with gluino is enhanced over
the one with neutralino. This leads to a larger ratio between
neutron EDM and electron EDM of $\gtrsim 10^{3}$. In typical
supersymmetric models with gaugino and sfermion masses of the same
order, this ratio is $\lesssim 10^{2}$~\cite{Abel:2005er}. This
seems to be a robust feature of models in which the gauginos are
suppressed relative to the squarks and the trilinears and the
trilinears are not proportional to Yukawas. Finally, it is easy to
see that the mercury EDM provides the most stringent limit on the
squark masses. For squark masses around $10$~TeV, the mercury EDM
will increase to $\sim 10^{-29}\,e\,{\rm cm}$, which could be
tested in the near future with better experimental precision.
Basically, we have found that the upper bounds on the EDMs in the
M-theory framework result from two of its generic features - large
scalar masses and trilinears (${\cal O}(10)$ TeV), and
CP-violating phases only in the trilinear couplings and those arising
only from the Yukawas.

\section{Higher-Order Corrections}\label{general}

In Section \ref{leading}, we found that there are no CP-violating
phases from supersymmetry breaking at leading order in the
framework of M-theory compactifications considered. It is
therefore important to check if corrections to the \kahler
potential and superpotential lead to further contributions to
CP-violating phases in the soft parameters and in turn to the
EDMs. Although the detailed form of possible corrections is not
known in M-theory, some general arguments can neverthless be made,
which strongly suggest that higher order corrections still
naturally suppress CP-violating phases.

The corrections to the soft parameters may arise in general from
corrections in the superpotential and the \kahler potential. In
the zero flux sector which we have considered, the superpotential
may only receive additional non-perturbative corrections from strong
gauge dynamics or from membrane instantons. As mentioned in section \ref{cpphases} and explained in Appendix (\ref{dynamic}), the dynamical alignment of
phases still works if these additional terms are subdominant compared to the first two terms. The subdominance of these additonal terms is required anyways to keep the results of the moduli stabilization mechanism intact and hence the consistency of the whole approach. Moreover, it also provides an elegant dynamical solution to the strong CP problem within string theory \cite{Acharya:2010zx}.

Th \kahler potential for the hidden sector comprising the moduli and hidden
matter fields on the other hand receives perturbative corrections such as terms with higher powers of $\phi$ as there is no non-renormalization theorem for the \kahler potential.
However, the field $\phi$ is composed of elementary quark fields
${\cal Q},{\cal \tilde{Q}}$ which are charged under the hidden
gauge group. Therefore, higher order corrections must be functions
of ${\cal Q}^{\dagger}{\cal Q}$ or ${\tilde {\cal
Q}}^{\dagger}\tilde {\cal Q}$ in order to be gauge invariant. When
written in terms of $\phi$, these corrections are always functions
of $\phi^{\dagger}\phi$. This structure is important for our claim
of small CP-violating phases since it does not introduce any new
phases in the soft parameters. In addition, the perturbative corrections
to the \kahler potential are always functions of $z_i+\bar z_i$, which do not lead to any CP violating phases in
the soft terms as argued in section \ref{cpphases}. The
dependence on $z_i+\bar z_i$ is a reflection of the shift (PQ) symmetry
of the axion $t_i \rightarrow t_i + \delta$, where $t_i=Im(z_i)$. This symmetry is only broken
by exponentially suppressed non-perturbative effects \cite{Silverstein:1996xp,Antoniadis:1997zt}. This implies that upto exponentially suppressed contributions, the corrections to the \kahler potential do not give rise to any CP violating phases in the soft terms. Thus, although it is very hard in
general to compute the form of corrections in M-theory, the result that the CP violating phases in soft parameters are highly suppressed should be quite robust as it only relies on symmetries.

\section{Conclusions}\label{conclude}

In this paper, we have discussed CP violation in theories arising
from fluxless M-theory compactifications with low energy
supersymmetry and all moduli stabilized. We have found that the
supersymmetry breaking dynamics is CP conserving (up to exponentially suppressed
effects). The gaugino masses, scalar masses, $\mu$ and $B\mu$ parameters are real
at the GUT scale and remain real at the Electroweak scale since RG evolution does not introduce
new CP-violtaing phases. However, the full trilinear couplings ${\hat A}_{\alpha\beta\gamma}=A_{\alpha\beta\gamma}Y_{\alpha\beta\gamma}$ manage to pick up
CP-violating phases from the Yukawa couplings as the trilinear
matrices are not proportional to the Yukawa matrices. In addition,
RG effects mix the phases between different flavors.

Given a model of Yukawas, therefore, one can estimate the effects of these
phases of Yukawas on the trilinear couplings, and therefore on the
EDMs. Since hierarchical Yukawa textures are well motivated within this framework,
we compute EDMs for such textures. The other relevant feature of the low energy theory is that
the scalar masses and trilinears are naturally of ${\cal
O}(m_{3/2})$ due to the absence of sequestering generically.
Morover, these scalar masses and trilinears are naturally of
${\cal O}(10$ TeV) since this is the natural scale of the
gravitino mass within the M-theory framework. These two features naturally
give rise to small CP-violating effects consistent with experimental limits.

We have estimated the electron, neutron and mercury electric
dipole moments utilizing the above features, and found that the
estimated upper bounds of the EDMs are all within current
experimental limits. The estimated upper bound for the mercury EDM
is close to the current experimental limit and could be probed in the
near future. A robust prediction of the framework is the existence
of a hierarchy of about three orders of magnitude between the
neutron and electron EDMs. This essentially results from the mass
hierarchy between gauginos and scalars as predicted within
M-theory \cite{Acharya:2007rc}, and provides an additional means
to test the framework.

It should be emphasized that our results
for EDMs are based on the result that the CP-violating phases are
entirely from the Yukawas, and therefore, any experimental result
which indicates other significant sources of phases would contradict
and rule out this approach. We also discuss effects of possible
corrections to the K\"{a}hler potential and superpotential, and
the generalization to other string compactifications.

Note that our results are largely independent of a full solution
to the flavor problem. Our results have been derived using the
fact that the squark matrices are approximately flavor diagonal at
low energies which is naturally predicted within the M-theory
framework, as explained in section \ref{kahlersuper}. Therefore,
any solution to the flavor problem consistent with the above
feature is consistent with our results.

The quark Yukawas give the CKM phases, and the lepton Yukawas the
PMNS phases. The latter can provide the phases needed for
baryogenesis via leptogenesis consistent with the above framework,
for example as described in \cite{Kumar:2008vs}. So even with no phases from
the soft supersymmetry breaking this framework can give a complete
description of all known CP violation.

\acknowledgments{We would like to thank Bobby Acharya,
Nima Arkani-Hamed, Konstantin Bobkov, Jacob Bourjaily, Lisa Everett,
Eric Kuflik, Arjun Menon, Brent
Nelson, Aaron Pierce and Liantao Wang for useful discussions. PK
would also like to thank the Michigan Center for Theoretical
Physics (MCTP) for their hospitality where part of the research
was conducted. The work of GLK and JS is supported in part by the
Department of Energy. The work of PK is supported by DOE under
contract no. DE-AC02- 05CH11231 and NSF grant PHY-04-57315.}

\appendix

\section{Dynamical Alignment of Phases in the Superpotential}\label{dynamic}

The dynamical alignment of phases is crucial for solving the SUSY
CP problem in the M-theory framework. It means that terms in the
moduli superpotential dynamically align to acquire the same phase in the vacuum (perhaps up to exponentially suppressed effects), implying that the superpotential in the vacuum becomes
\emph{real} after rotating away the unphysical overall phase.

It was shown in \cite{Acharya:2007rc, Acharya:2008hi} that with just two non-perturbative contributions (``double condensate terms") in the superpotential, all $N$ moduli, the meson field, and one axion can be fixed in the supergravity regime with both terms in the superpotential acquiring a common phase in the vacuum. At this level, however, all but one axions remain unfixed. This is because the supergravity scalar potential only depends on one linear-combination of axions, while it depends on all the moduli (through the K\"{a}hler derivative). It is important to stabilize the remaining $N$ axions in a way such that the superpotential becomes (dominantly) real in the vacuum.

Precisely such a mechanism to stabilize the axions has been recently studied in \cite{Acharya:2010zx}. There, it was shown that higher order terms in the superpotential, which depend on the remaining linear combinations of axions, can stabilize these axions. The superpotential is then given by:\be\label{susimp}
W=A_1\phi^ae^{-b_1f}+A_2e^{-b_2f}+ \sum_{k>2}A_k e^{-b_k f^k}\,,
\ee

where we have assumed that the subdominant terms in (\ref{susimp}) arise from string instantons or gaugino condensates in pure Super Yang-Mills hidden sectors. It is possible to include matter in the hidden sectors as well, but that will not change the qualitative results obtained. In order to keep the original moduli stabilization mechanism and the resulting analysis of \cite{Acharya:2007rc, Acharya:2008hi} intact, the ``double condensate" terms must be parametrically larger than these higher order terms. This can be naturally obtained when $b_1 \sim b_2 < b_k,\, k>2$. It turns out that this will also make the superpotential real in the vacuum
(up to exponentially suppressed effects), as shown below.

It is easy to see that in the ${\cal N}=1$ supergravity potential, the dominant terms arise from the first two terms in (\ref{susimp}). Following \cite{Acharya:2007rc}, the dependence of the dominant potential on the axions is given by:
\ba V_{dom}(t_i) &=& \frac{1}{{\cal V}^3}[\sum_{\alpha=1}^2(\sum_{i=1}^N\,V^{(1)}_{\alpha\,i}+
V^{(2)}_{\alpha})\,e^{-2b_{\alpha}\mathrm{Re}(f)}+\nonumber\\ && \sum_{i=1}^N\,(V^{(3)}_i + V^{(4)})
\,e^{-(b_{1}+b_2)\mathrm{Re}(f)}\times\nonumber\\ &&  \cos{(\chi_1-\chi_2)]}\ea where $\chi_i\equiv b_i\vec{N}^i\cdot\vec{t}+\delta_{i1}\,a\theta$, $\mathrm{Im}(f)\equiv \vec{N}\cdot\vec{t}$, $\phi=\phi_0\,e^{i\theta}$, and
$V^{(1)}_{\alpha\,i}$, $V^{(2)}_{\alpha}$, $V^{(3)}_i$, $V^{(4)}$ are \emph{positive} coefficients independent of the axions. Thus, the axion combination $\chi_1-\chi_2$ is minimized at \be\label{co}\cos{(\chi_1-\chi_2)}=\cos[(b_1-b_2)\vec{N}\cdot\vec{t}+a\theta]=-1\ee while the remaining axion combinations stay unfixed.

Now, including other terms in the superpotential such that the ``double condensate" approximation is valid, one finds that the remaining $N$ axions are stabilized by the next $N$ largest terms in the scalar potential depending on those $N$ linearly independent combinations of axions. These terms are generated by the product of one of the two dominant terms and one of the subdominant terms. To compute the stabilized values of the axion combinations $\chi_1-\chi_k;\;k>2$ it is useful to write the effective potential for light axions after integrating out the moduli, meson field and one axion combination which receive a mass of ${\cal O}(m_{3/2})$. This is given by \cite{Acharya:2010zx}: \ba\label{veff}
V_{eff}(\chi_i)&\approx& V_0-m_{3/2}e^{K/2}\sum_{k=3}^{N+2}D_ke^{-b_kV_k}\cos(\chi_1-\chi_k)
\nonumber\\&\forall k:&\,b_kV_k<b_{k+1}V_{k+1}\,.
\ea where $D_k$ are \cal{O}(1) coefficients. This implies that the $N$ axion combinations $\chi_1-\chi_k$ are all stabilized such that: \ba\label{others} \cos{(\chi_1-\chi_k)}=\pm 1;\,\,\,k=3,..,N+2\ea depending on whether $D_k$ is negative or positive respectively. In terms of $\chi_i$, the full superpotential can thus be written as: \ba \label{fullW}W&=&e^{i\chi_1}(|W_1|+|W_2|e^{-i(\chi_1-\chi_2)}+\sum_{k=3}^{N+2}|W_k|e^{-i(\chi_{1}-\chi_k)}+..)\nonumber\\&=&e^{i\chi_1}\left(|W_1|-|W_2|\pm \sum_{k=3}^{N+2}|W_k|\right)+...\nonumber\ea which is (dominantly) real up to one  overall phase. The  phases of other possible terms present in the superpotential (denoted by ... above) will be completely determined by the stabilized axions above, and may be different from 0 or $\pi$. However, since these terms are exponentially suppressed relative to the dominant terms, one finds that after rotating away the unphysical overall phase the superpotential is real in the vacuum up to exponentially suppressed effects.

Note that for the above mechanism to work it is crucial that the terms in the superpotential which stabilize all moduli and one combination of axions, are dominant compared to the remaining terms.
In the case studied in this paper this amounts to having the first two terms in the superpotential dominant compared to other ones. If many or all terms are comparable to each other, the analysis becomes more complicated and the axion combinations $\chi_1-\chi_k;\,k=3,4,..$ are then generically stabilized at a non-trivial value other than 0 or $\pi$ implying that the superpotential is not real in the vacuum.

Another consequence of the above result which has been studied in \cite{Acharya:2010zx} in detail, is that these light axion mass eigenstates are exponentially lighter compared to the moduli, meson and heavy axion which are ${\cal O}(m_{3/2})$. This allows for a beautiful dynamical solution to the strong CP problem within string theory and explicitly realizes the string axiverse scenario considered in \cite{Arvanitaki:2009fg}. It is remarkable that the mechanism which stabilizes the axions in the framework considered in \cite{Acharya:2010zx} and solves the strong CP problem automatically stabilizes them in such a way that the superpotential is (dominantly) real in the vacuum.

The above result also applies to other classes of string compactifications. For example, it was shown in \cite{Bobkov:2010rf} that within certain classes of Type IIB flux compactifications, all k\"{a}hler moduli and one of their axion partners can be stabilized with a superpotential with a constant term and just one non-perturbative contribution which depends on a linear combination of all moduli, i.e. for \ba W &=& W_0 + A\,e^{-b\,f};\nonumber\\ f&=&\sum_i^N N^iT_i\ea All but one axions remain massless at this level. If these axions are stabilized by subdominant terms in the superpotential, this would give rise to the same conclusion in these compactifications as well.

\section{The leading one-loop contributions to
EDM}\label{appendix:one-loop}

The fermion EDMs can be generated at one-loop in supersymmetric
models with CP-violating phases in the soft supersymmetry breaking
sector. Within the framework considered in this paper, the
CP-violating phases only reside in the trilinear terms and
therefore appear in the mass mixing terms of the left- and
right-handed sfermions. Therefore, the main contribution to the
quark EDM and CEDM comes from diagrams involving gluinos because
of the large gauge coupling. For the electron EDM, the dominant
contribution comes from the diagram involving neutralinos. This is
because the diagrams with charginos in the loop require
CP-violating phases in the chargino sector which do not arise
within the M-theory framework considered.

Let us first consider the diagrams contributing to quark EDM and
CEDM with gluino running in the loop
\begin{eqnarray}
d_{q}^{\tilde{g}(E)}& = & \frac{-2e\alpha_{s}}{3\pi}\sum_{k=1}^{2}
{\rm
Im}(\Gamma_{q}^{1k})\frac{m_{\tilde{g}}}{m_{{\tilde{q}}_{k}}^{2}}
Q_{\tilde{q}}B\left(\frac{m_{\tilde{g}}^{2}}{m_{{\tilde{q}}_{k}}^{2}}\right)\\
d_{q}^{\tilde{g}(C)} & = &
\frac{g_{s}\alpha_{s}}{4\pi}\sum_{k=1}^{2}{\rm
Im}(\Gamma_{q}^{1k})\frac{m_{\tilde{g}}}{m_{{\tilde{q}}_{k}}^{2}}
C\left(\frac{m_{\tilde{g}}^{2}}{m_{{\tilde{q}}_{k}}^{2}}\right)
\end{eqnarray}
where $\Gamma_{q}^{1k}=D_{q2k}D_{q1k}^{*}$ and $D_{q}$ is the
$2\times2$ matrix which diagonalizes the squark mass matrix
$m_{\tilde{q}}^{2}$
\begin{equation}%
D_{q}^{\dagger}m_{q}^{2}D_{q}={\rm
Diag}(m_{\tilde{q}1}^{2},m_{\tilde{q}2}^{2}).
\end{equation}%
More explicitly
\begin{eqnarray}%
{\tilde{q}}_{L} & = & D_{q11}\tilde{q}_{1}+D_{q12}\tilde{q}_{2}\nonumber \\
{\tilde{q}}_{R} & = &
D_{q21}\tilde{q}_{1}+D_{q22}\tilde{q}_{2}.
\end{eqnarray}%
Here $B(r)$ and $C(r)$ are loop functions defined as:
\begin{eqnarray}%
B(r) & = & \frac{1}{2(r-1)^{2}}\left(1+r+\frac{2r\ln(r)}{1-r}\right)\nonumber \\
C(r) & = &
\frac{1}{6(r-1)^{2}}\left(10r-26+\frac{2r\ln(r)}{1-r}-\frac{18\ln(r)}{1-r}\right)\nonumber
\end{eqnarray}%
In the above equations, we assume no flavor mixing in the squark
mass matrices as argued in the main body of the paper. Using the fact that ${\rm Im}(\Gamma_{q}^{11})=-{\rm
Im}(\Gamma_{q}^{12})$, we have:
\begin{eqnarray}
d_{q}^{\tilde{g}(E)}  \approx
\frac{-2e\alpha_{s}Q_{\tilde{q}}}{3\pi}\frac{{\rm
Im}({m_{\tilde{q}}^{2}})_{LR}}{m_{\tilde{g}}^{3}}
r^{2}\left(B(r)+r\, B'(r)\right)
\end{eqnarray}
Similarly
\begin{eqnarray}
d_{q}^{\tilde{g}(C)}  \approx
\frac{g_{s}\alpha_{s}}{4\pi}\frac{{\rm
Im}({m_{\tilde{q}}^{2}})_{LR}}{m_{\tilde{g}}^{3}}
r^{2}\left(C(r)+r\, C'(r)\right)
\end{eqnarray}
In the calculation above, we assume the mass splitting of squarks
is small compared to the squark mass. This is usually true since
we are only interested in the up and down squarks. When
$r=m_{\tilde{g}}^{2}/M_{\tilde{q}}^{2}\ll1$,
one finds that $C(r)\gg A(r),B(r)$. %$B'(r)\sim 0.5$ and $C'(r)\sim -3/x$.
It is easy to see that $d_{q}^{\tilde{g}(C)}\gg
d_{q}^{\tilde{g}(E)}$. For other diagrams which involve neutralino
and charginos, the structure is very similar. However, they are
much smaller than $d_{q}^{\tilde{g}}$ and can be neglected.

Now let us turn to the one-loop diagrams contributing to the
electron EDM
\begin{align}
d_{e}^{\tilde{\chi}^{+}} &=
\frac{e\alpha_{em}}{4\pi\sin^{2}\theta_{W}}\sum_{k=1}^{2}{\rm
Im}(\Gamma_{ei})\frac{m_{\tilde{\chi}^{+}}}{m_{\tilde{\nu}}^{2}}
A\left(\frac{m_{\tilde{\chi}^{+}}^{2}}{m_{\tilde{\nu}}^{2}}\right)\label{chargino}\\
d_{e}^{\tilde{\chi}^{0}} &=
\frac{e\alpha_{em}}{4\pi\sin^{2}\theta_{W}}\sum_{k,i=1,1}^{2,4}{\rm
Im}(\eta_{eik})\frac{m_{\tilde{\chi}^{0}}}{m_{{\tilde{e}}_{k}}^{2}}B
\left(\frac{m_{\tilde{\chi}^{0}}^{2}}{m_{{\tilde{e}}_{k}}^{2}}\right)\label{neutralino}
\end{align}
where $\Gamma_{ei}=U_{i2}^{*}V_{i1}^{*}$, and
\begin{eqnarray}
\eta_{eik}= \Big[-\sqrt{2}\left\{ \tan\theta_{W}(Q_{e}-T_{3e})X_{1i}+T_{3e}X_{2i}\right\} D_{e1k}^{*}\nonumber\\
 + \kappa_{e}X_{bi}D_{e2k}^{*}\Big]\left(\sqrt{2}\tan\theta_{W}Q_{e}X_{1i}D_{e2k}-\kappa_{e}X_{bi}D_{e1k}\right)\nonumber
\end{eqnarray}
Here we have
\begin{equation}
\kappa_{e}=\frac{m_{e}}{\sqrt{2}m_{W}\cos\beta}
\end{equation}
The loop function $A(r)$ is given by
\begin{equation}
A(r)=\frac{1}{2(1-r)^{2}}\left(3-r+\frac{2\ln(r)}{1-r}\right)
\end{equation}
In the above equations, $U(V)$, $X$ and $D_{e}$ are the
conventional chargino, neutralino and selectron mixing matrices.
It is easy to see that the chargino diagram do not contribute to
the electron EDM in the framework considered, since there is no
CP-violating phases in the chargino sector. In the absence of the
neutralino mixing, the expression of $d_{e}^{\tilde{\chi}_{0}}$
can be significantly simplified
\begin{eqnarray}\label{de-neutralino}
d_{e}^{E} \approx
\frac{e\alpha_{em}}{4\pi\cos^{2}\theta_{W}}\frac{{\rm
Im}(m_{\tilde{e}}^{2})_{LR}}{m_{\tilde{B}}^{3}}\;
r_{1}^{2}\left[B(r_{1})+r_{1}B'(r_{1})\right]
\end{eqnarray}
where $r_{1}=m_{\tilde{B}}^{2}/m_{\tilde{e}}^{2}$ with
%$r_{2}=m_{\tilde{H}_{d}}^{2}/m_{\tilde{e}}^{2}$, and
$m_{\tilde{e}}$ denoting the average mass of the selectrons. In
the above result, the higgsino contribution is neglected since it
is suppressed by the small $Y_e^2$.

%******************************************************************************************

\section{Barr-Zee diagram}\label{appendix:barr-zee}
As we have discussed in subsection \ref{sec:two-loop}, we are
concerned with the Barr-Zee diagram with the third generation
squarks, i.e. $\tilde t$ and $\tilde b$, running in the inner
loop. Here we give the detailed derivation of
Eq.~(\ref{Eq:Barr-Zee1}) and (\ref{Eq:Barr-Zee2}). We start with
the general results of EDM and CEDM for the Barr-Zee diagram
~\cite{Chang:1998uc}
\begin{eqnarray}%
d_{f}^{E}&=&
Q_{f}\frac{3e\alpha_{em}}{32\pi^{3}}\frac{R_{f}m_{f}}{M_{A}^{2}}
\sum_{q=t,b}\xi_{q}Q_{q}^{2}\left[F(r_1)-F(r_2)\right]\nonumber \\
d_{f}^{C} & =&
\frac{g_{s}\alpha_{s}}{64\pi^{3}}\frac{R_{f}m_{f}}{M_{A}^{2}}\sum_{q=t,b}\xi_{q}\left[F(r_1)-F(r_2)\right]
\label{two-loop-1}
\end{eqnarray}
where $M_A$ is the mass of pseudoscalar higgs $A_0$,
$r_{1,2}=m_{{\tilde q}_{1,2}}^2/M_A^2$,
$R_{f}=\cot\beta$($\tan\beta$) for $I_{3}=1/2$($-1/2$) and $F(z)$
is the two-loop function
\begin{equation}
F(z)=\int_{0}^{1}dx\frac{x(1-x)}{z-x(1-x)}\ln\left[\frac{x(1-x)}{z}\right].
\end{equation}
The CP-violating couplings are given by \begin{eqnarray}
\xi_{t} & = & -\frac{\sin2\theta_{\tilde{t}}m_{t}{\rm Im}(\mu e^{i\delta_{t}})}{2v^{2}\sin^{2}\beta}\nonumber \\
\xi_{b} & = & -\frac{\sin2\theta_{\tilde{b}}m_{b}{\rm
Im}(A_{b}e^{-i\delta_{b}})}{2v^{2}\sin\beta\cos\beta}
\label{xi-tb}
\end{eqnarray}
where $\theta_{\tilde{t},\tilde{b}}$ are the stop and sbottom
mixing angles, and $\delta_{q}={\rm Arg}(A_{q}+R_{q}\mu^{*})$. The
mixing angle of the squark sector is given by%
\begin{eqnarray}%
\tan(2\theta_{q}) & = & -\frac{2m_{q}|\mu R_{q}+A_{q}^{*}|}{M_{\tilde{Q}}^{2}-M_{\tilde{q}}^{2}+\cos2\beta M_{Z}^{2}(T_{z}^{q}-2e_{q}s_{w}^{2})}\nonumber \\
 & \approx & -\frac{2m_{q}|\mu R_{q}+A_{q}^{*}|}{M_{\tilde{Q}}^{2}-M_{\tilde{q}}^{2}}\label{tan2thetaq}
\end{eqnarray}%
Therefore, Eq.~(\ref{xi-tb}) becomes%
\begin{eqnarray}%
\xi_{t} & \approx & \frac{y_{t}^{2}|A_{t}^{*}+\mu\cot\beta|{\rm Im}(\mu e^{i\delta_{t}})}{M_{\tilde{Q}}^{2}-M_{\tilde{t}}^{2}}\nonumber \\
\xi_{b} & \approx &
\cot\beta\frac{y_{b}^{2}|A_{b}^{*}+\mu\tan\beta|{\rm
Im}(A_{b}e^{-i\delta_{b}})}{M_{\tilde{Q}}^{2}-M_{\tilde{b}}^{2}}\label{xi-tb-1}%
\end{eqnarray}%
Using Eq.~(\ref{xi-tb}) and (\ref{tan2thetaq}), we can rewrite
Eq.~(\ref{two-loop-1}) as%
\begin{eqnarray}\label{bar-zee}
d_{f}^{E}&\approx &
Q_{f}\frac{3e\alpha_{em}}{32\pi^{3}}\frac{R_{f}m_{f}}{M_{A}^{4}}
{\rm Im}\bigg[\frac{4y_{t}^{2}}{9}\mu(A_{t}+\mu^{*}\cot\beta)F'\left(r_{1}\right)\nonumber\\
& + & \frac{y_{b}^{2}}{9}A_{b}(A_{b}^{*}+\mu\tan\beta)\cot\beta F'\left(r_{2}\right)\bigg]\nonumber\\
d_{f}^{C} & \approx & \frac{g_{s}\alpha_{s}}{64\pi^{3}}
\frac{R_{f}m_{f}}{M_{A}^{4}} {\rm Im}\bigg[ y_{t}^{2}\mu(A_{t}+\mu^{*}\cot\beta)F'\left(r_{1}\right)\nonumber\\
& + & y_{b}^{2}A_{b}(A_{b}^{*}+\mu\tan\beta) \cot\beta
F'\left(r_{2}\right)\bigg]
\end{eqnarray}
where $r_{1}\equiv m_{\tilde{t}}^{2}/M_{A}^{2}$ and $r_{2}\equiv
m_{\tilde{b}}^{2}/M_{A}^{2}$ with $m_{\tilde{t},\tilde{b}}$ being
the average masses of the stops and sbottoms respectively.

%*****************************************************************************************

%\section{Purely Gluonic Dimension Six Operators}\label{dim6}

%The gluonic dimension $6$ operator can arise from the two-loop
%diagram involving gluino, squark and quark in the loop as seen in
%Fig.~\ref{Weinberg}. It is first calculated in \cite{Dai:1990xh},
%which is
%\begin{eqnarray}
%d^{G} & =&
%-3\alpha_{s}\left(\frac{g_{s}}{4\pi}\right)^{3}\frac{{\rm
%Im}(m_{\tilde{q}}^{2})_{LR}}{m_{t}m_{\tilde{g}}^{3}}z_{t}H(z_{1},z_{2},z_{t})\nonumber\\
%& + &(t\rightarrow b),
%\end{eqnarray}
%where
%$z_{i}=\left(\frac{m_{{\tilde{t}}_{i}}}{m_{\tilde{g}}}\right)^{2}$,
%$z_{t}=\left(\frac{m_{t}}{m_{\tilde{g}}}\right)^{2}$. The two-loop
%function $H(z_{1},z_{2},z_{t})$ is given in \cite{Dai:1990xh}. For
%the case that only trilinear couplings contain phases, the above
%equation can be written as

%One can see that with a large non-universal A-term, this contribution could be dominant for the quark EDM and so neutron EDM.
%As an example, we take one of the $G_2$-MSSM benchmark point with $m_{\tilde t_1}=3.85$ TeV, $m_{\tilde t_2}=12.4$ TeV, $m_{\tilde g}=573.5$ GeV
%and $A_t=2.9$ TeV. The contribution to neutron EDM from stop-top loop is $d_n^{G(\tilde t)}\approx 1.8\times 10^{-27}$ e cm, which is only an order of
%magnitude smaller than the current experimental bound.

\bibliography{susy-cp}

\end{document}